\documentclass[ecta,nameyear,final]{econsocart}
\RequirePackage[colorlinks,citecolor=blue,linkcolor=blue,urlcolor=blue,pagebackref]{hyperref}

\startlocaldefs
\theoremstyle{plain}

\theoremstyle{remark}

\usepackage{graphicx,subcaption}
\usepackage{tikz}
\usetikzlibrary{shapes,decorations,arrows,calc,arrows.meta,fit,positioning}
\tikzset{
    -Latex,auto,node distance =1 cm and 1 cm,semithick,
    state/.style ={ellipse, draw, minimum width = 0.7 cm},
    point/.style = {circle, draw, inner sep=0.04cm,fill,node contents={}},
    bidirected/.style={Latex-Latex,dashed},
    el/.style = {inner sep=2pt, align=left, sloped}
}
\newcommand\bigzero{\makebox{\text{\large\bfseries 0}}}\usepackage{amsmath}
\usepackage{nicematrix}
\usepackage{mathtools}  
\DeclarePairedDelimiterX{\infdivx}[2]{(}{)}{%
  #1\;\delimsize\|\;#2%
}

\usepackage{siunitx}
\usepackage{enumitem}
\usepackage{tcolorbox}
\endlocaldefs

\begin{document}
\begin{frontmatter}

    \textbf{Treatment effects without multicollinearity? Temporal order and the Gram-Schmidt process in causal inference.}    
    \runtitle{\textsc{Cross and Buccola}} \newline

    \small 
    
    \textbf{Robin M. Cross and Steven T. Buccola} \\

    \scriptsize

    Department of Applied Economics, Oregon State University, Corvallis, Oregon, USA \newline

    \textbf{Correspondence} 

    Robin M. Cross, 221B Ballard Hall, Corvallis, Oregon 97331. 
    
    Email: robin.cross@oregonstate.edu \\

    \support{We are grateful to Anthony Nearman, Karen Rennick, Nathalie Steinhaur, and Dennis vanEnglesdorp for the initial motivation to solve the multicollinearity problem, W. Jason Beasley and Aaron Watt for experimental implementation, J. Aislinn Bohren and Peter Hull for comments on discrimination decomposition in survey data, Jennifer Alix-Garcia, David Kling, and David Lewis for input on related research and project scope, and Guido Imbens for encouragement to address temporal order. Juan Carlos L\'{o}pez-Morate provided an excellent proof for the lower standard errors under included-irrelevant variables property Theorem 2(G).}

     \begin{tcolorbox}[colframe=white, colback=gray!20, sharp corners=all]
         
        \footnotesize \textbf{Summary} \\
        
        This paper incorporates information about the temporal order of regressors to estimate orthogonal and economically interpretable regression coefficients. We establish new finite sample properties for the Gram-Schmidt orthogonalization process. Coefficients are unbiased and stable with lower standard errors than those from Ordinary Least Squares. We provide conditions under which coefficients represent average total treatment effects on the treated and extend the model to groups of ordered and simultaneous regressors. Finally, we reanalyze two studies that controlled for temporally ordered and collinear characteristics, including race, education, and income. The new approach expands Bohren \textit{et al.}'s decomposition of systemic discrimination into channel-specific effects and improves significance levels.\newline
        
        \scriptsize  \textbf{KEYWORDS} \\
        
        collinear regressors, recursive Laplace orthogonalization, QR decomposition, temporal order
    
    \end{tcolorbox}
    
\end{frontmatter}

\section{INTRODUCTION}\label{s1} 

\noindent Multicollinearity, or correlation among regressors, has posed a problem for statistical analysis since the introduction of Ordinary Least Squares (OLS) in \citeyear{legendre}. Inflated standard errors depress statistical significance, and negative coefficient covariance renders models sensitive to small changes in regressor selection or functional form. A growing area of concern has been the emergence of \textit{p}-hacking, the selection of regressors or functional forms to inflate statistical significance \citep{leamer, gelman}. At the same time, rapid improvements in computing power and the emergence of machine-learning have increased the number of closely related regressors that can be analyzed in a single model \citep{baiardi}. The statistical instability induced by even modest levels of multicollinearity in the data exacerbates its influence on reported findings, as strategic or arbitrary inclusion of even a single regressor can radically vary the conclusions. All this raises the stakes for discovering an alternative to OLS that will be both rationally interpretable and robust in the face of multicollinearity.

Various statistical treatments for multicollinearity have been developed, each coming at some cost in terms of either coefficient bias or interpretability. Step-wise methods \citep{hocking} and discretionary elimination of correlated regressors introduce omitted-variable bias. The advantage of retaining all regressors via ridge regression \citep{hoerl} comes at the cost of systematic downward bias. 

Orthogonalization is a popular machine-learning solution for computation, forecasting, and signal processing wherever directly interpretable coefficients are not required \citep{ despois}. \citet{laplace} introduced his orthogonalization approach to compute \citeauthor{legendre}’s newly popularized (\citeyear{legendre}) least-squares regression, free of \citeauthor{gauss}'s (\citeyear{gauss}) computationally more expensive normal equations \citep[see][]{stigler}. Laplace's method was independently discovered by \citet{gram}, \citet{schmidt}, and \citet{iwasawa}, and coined the Modified Gram-Schmidt Process by \citet{wong} \citep[see][]{farebrother}.\footnote{\citet{langou} provides an English translation of Laplace's original \citeyear{laplace} manuscript.} The Gram-Schmidt is a special case of upper-triangular matrix decomposition, referred to by \citet{francis} as QR decomposition, which preserves all information in the original data set, a so-called lossless process. A long list of orthogonal models have followed Laplace, notably \citeauthor{pearson}'s (\citeyear{pearson})  lossless Principal Component Analysis (PCA) for covariate-importance ranking and \citeauthor{golub}'s (\citeyear{golub})  Singular Value Decomposition (SVD) eigenvalue-based method for data compression. 

Research interest in multicollinearity has waned in recent decades due in part to a growing consensus that growing data sizes will ameliorate multicollinearity's impact \citep{greene} and partly from the unreliability of existing diagnostic approaches like pairwise correlation and the Variance Inflation Factor (VIF) \citep{kalnins23}. In addition, multicollinearity is commonly perceived to be a statistical rather than economic problem \citep{enikolopov}. Unfortunately, statistical confidence reported for the incorrect sign (Type I error), so-called sign-switching, increases with sample size \citep{kalnins18}. We will explore how multicollinearity's statistical symptoms may arise from omitting readily available economic information heretofore overlooked -- the temporal order of the regressors.

This paper makes four contributions. First, we derive the finite sample properties of the Gram-Schmidt least-squares model. We show how the model preserves all regressor information and that its coefficients are unbiased and stable, with standard errors lower than in OLS. Second, we explore conditions under which the model returns average treatment effects when treatment responses are heterogeneous. We show that for late treatments - those applied after all other personal characteristics have been determined - the estimate is identical to OLS, and the average treatment effect bias can be estimated and corrected with existing approaches \citep{sloczynski}. For early treatments, applied before other characteristics and associations are formed, the model returns an unbiased estimate of the average total treatment effect on the treated (ATTT). Third, we extend the model to groups of simultaneous regressors common in economic data sets, expanding the model's use beyond strictly temporally distinct regressors.\footnote{Here, we refer to regressors that are simultaneous to one another rather than to the regressand or error term as are usually considered in simultaneous and endogenous models.} Finally, we apply our model to \citeauthor{sloczynski}'s (\citeyear{sloczynski}) decomposition of \citeauthor{angrist09}'s (\citeyear{angrist09}) analysis of job program effectiveness, and \citeauthor{lubotsky}'s (\citeyear{lubotsky}) assessment of contributors to child reading-and-comprehension outcomes. Both studies controlled for several collinear and temporally-ordered individual characteristics. Our analysis extends earlier findings to the \citet{bohren} decomposition, which breaks total discrimination into its direct and systemic elements. Our model expands this decomposition, dividing systemic discrimination further into such channel-specific mechanisms as education, income, and household formation.

Our interpretability result is motivated by earlier research in interdependent dynamic systems, first explored in the 1920s by economists interested in describing and forecasting simultaneous supply and demand relations. There, system identification was a primary focus. Early strategies included the instantaneous equilibrium condition, lagged endogenous variables, causal chains, and instrumental variables. \citet{schultz} introduced the cobweb model, extended by \citet{wright25, wright34} to path analysis and instrumentation. Wold generalized these causal chain and dynamic system approaches to the Linear Systems of Equations Model (LSEM) (\citeyear{wold51, wold60}), showing in a time-series framework that model coefficients were asymptotically consistent (\citeyear{wold63}). \citet{goldberger} explored systems with unobservable regressors and early latent factor models. \citet{alwin} first calculated indirect effects and their confidence intervals by using a product-of-coefficients method.

\citet{rubin} refocused causal identification away from LSEM's lagged variables and instantaneous equilibria toward the random control trial (RCT) strategies inspired by \citet{neyman}, the so-called Rubin causal model \citep[see][]{holland}, or Potential Outcomes approach \citep[see][]{imbens20}. \citet{imbens94} and \citet{card} extended these strategies greatly, particularly the use of instrumental variables, difference-in-difference, and shape restrictions, frequently considering models with potential endogeneity from omitted variables. The resulting Causal Inference (CI) framework is used widely across academic disciplines.

Parallel to these efforts, \citet{pearl00} expanded LSEM to the probabilistic, graph-centric Causal Mediation (CM) framework, of growing popularity in the biological, computer, and social sciences outside economics. \citeauthor{pearl00} showed a regressor's total effect to be a partial derivative of an independent variable with respect to the regressor's path through the CM system \citep{hunermund}, reformulating \citeauthor{wright34}'s (\citeyear{wright34}) method of path coefficients and extending \citeauthor{wold63}'s chain principle (\citeyear{wold63}) beyond a time series setting. \citeauthor{pearl00}  recast \citeauthor{wold63}'s conditions in terms of a probabilistic Directed Acyclic Graph (DAG) or Bayesian Network. One-way causality is achieved either when regressors are temporally separated from one another by the natural occurrence of the data or by experimental design, referred to as Sequential Ignorability. \citet[p. 418]{pearl01} shows identification in this system is equivalent to the Gauss-Markov assumptions under the additional conditions of independent residual vectors and freedom from heterogeneity induced by unobservable regressors.\footnote{We show later how independent residuals and non-heterogeneity from unobservable regressors follow directly from the Gauss-Markov zero-conditional-mean condition and linearity assumptions, respectively.} 

Simultaneous regressors remain unidentified in both the LSEM and CM frameworks.\footnote{\citet[Proposition 9]{wold63} shows that the reduced-form parameters of an interdependent (simultaneous) system are not identified.} We utilize this graphical representation to motivate the linkage between temporal order and direct and total effects and provide properties sufficient for equivalence between the two frameworks. That DAGs follow from the LSEM framework is well documented, and we will not provide further equivalence conditions here.

Temporal order is by no means the only path to separability among regressors. It may be achieved either partially or fully by experimental design \citep{imbens18} or awareness of the underlying economic or other known causal mechanisms \citep{borusyak}. Temporal order is readily observable from data-collection frequency and timing. The passage of time precludes feedback (simultaneity) among regressors in a way that other data-collection or generating processes sometimes lack. In certain cases regressors may be simultaneously determined when measured over a given time scale, in monthly average expenditures and prices for instance, but well-ordered over smaller time scales such as daily transactions with posted prices.\footnote{\citet[p. 37]{lewis-beck} provide a helpful discussion of the source of simultaneity posed by \citet{strotz} and developed by \citet{fisher} and \citet{johnston}. They suggest that, by nature, regressors tend to arise sporadically rather than all-at-once. Even when feedback occurs between two particular regressors, it is usually by way of a succession of stimuli and responses. In brief, regressor simultaneity arises in naturally occurring data whenever its collection intervals span regressor creation and feedback response activities.} Finally, temporal order removes multicollinearity but does not preclude endogeneity resulting from unobserved regressors, a phenomenon explored by \citet{imbens94}. The relationship between temporal and unobserved regressors is being explored in a companion paper.

Linearity will be an important motivating assumption throughout this paper. It is restrictive and no longer explicitly imposed by many relevant CM and CI models \citep{hunermund, pearl01}. Despite this, linearity provides a tractable approach to temporal order and an accessible alternative to OLS, which remains a persistent tool for applied research \citep{imbens15} with growing interest from machine-learning CM approaches \citep{kumor}. 

This paper then proceeds as follows. The next section introduces the Gram-Schmidt process for a simplified data set of two recursive regressors. Section 3 reviews the LSEM and recursive DAG frameworks and provides conditions under which Gram-Schmidt coefficients are equivalent in expected value to a recursive LSEM representation. Section 4 derives finite-sample estimation properties for our proposed model. Section 5 relaxes homogeneity and provides the conditions for decomposing the Gram-Schmidt coefficients into average total treatment effects for both early and late-occurring treatments. Section 6 extends the method to a mixed data set containing simultaneous as well as recursive regressors and derives the extended estimation properties. Section 7 illustrates the new model by replicating and expanding the work of \citet{angrist09} and \citet{lubotsky}, who consider the effects of collinear and temporally-ordered family and individual characteristics on earnings and childhood reading and comprehension. Section 9 concludes.

\section{THE GRAM-SCHMIDT PROCESS}
\noindent The Gram-Schmidt process begins with a data set $X$ consisting of $K$ real-valued variables arranged in random order. The second variable is regressed on the first and replaced with the resulting residual; then, the regression coefficient is saved. This process is repeated, each variable regressed successively on all prior and then replaced by the resulting residual, saving the coefficients. The outcome is an upper-triangular matrix of estimation coefficients $C$ and a new variable matrix $U$ constituting the orthogonal basis of the original data set, essentially preserving all variable information but in orthogonal form. The final step is to divide each residual by its standard deviation, creating the regressor set's orthonormal basis. The latter step will be excluded below, although it is sometimes useful, and such coefficients can be interpreted in terms of standard deviation units.

The process can be represented by a set of line-by-line OLS regression equations, with each residual replaced by the prior equation's residual. The simple recursive system represented in Figure 1(a) can be represented as follows:
\newline
\setlength{\abovedisplayskip}{0pt}
\setlength{\belowdisplayshortskip}{0pt}
\begin{align}
\label{eqn:laplace0}
x &= u_x, \\
d &= c_{xd} u_x + u_d, \\
\label{eqn:laplace}
y &= c_{xy} u_{x} + c_{dy} u_d + u_y,
\end{align} 

\noindent where $c_{xd}$, $c_{xy}$, and $c_{dy}$ are estimated coefficients, $x$, $d$, and $y$ are data vectors, and $u_x$, $u_y$, and $u_d$ the residuals.  The coefficients can be specified as the inner-product ratios $c_{xd} = x'd / x'x$, where $'$ is the transpose operator.\footnote{For comparability with OLS, the algorithm is presented here in the inner-product form suggested by \citet{longley}, under which Laplace's computational accuracy remains identical to that of the modified Gram-Schmidt. See \citet{farebrother}.}

In matrix form, the system including $y$ can be written as $X=U(C+I)$, where the $(K+1) \times N$ matrix $X = [x \; d \; y]$ is decomposed into three components: (i) the $(K+1) \times N$ matrix $U$, containing orthogonal residuals vectors $[u_x \; u_d \; u_y]$; (ii) the identity matrix $I$; and (iii) the upper-triangular $K \times K$ matrix 
\setlength{\abovedisplayskip}{10pt}
\setlength{\belowdisplayshortskip}{0pt}
\begin{equation}
\label{eqn:matrixC}
C=
\begin{bmatrix}
0  & \hspace{0.3cm} c_{xd} & \hspace{0.3cm} c_{xy} \\
0 & \hspace{0.3cm} 0 & \hspace{0.3cm} c_{dy} \\
0 & \hspace{0.3cm} 0 & \hspace{0.3cm} 0 
\end{bmatrix} .
\end{equation}

This procedure produces a convenient orthogonal data set $U$. But how can coefficient matrix $C$ be interpreted? To answer this, we next explore the LSEM and recursive DAG models.

\section{ECONOMIC INTERPRETATION AND EQUIVALENCE}

\noindent DAGs are one approach to formulating a structural system of equations, deriving the reduced form, checking identification, and recovering the parameters. Three types of parameters are recovered in such systems: (i) direct effects, where a regressor acts, as is the case with OLS, directly on the dependent variable, holding other regressors constant; (ii) indirect effects, when a first regressor acts indirectly on the dependent variable by influencing a second regressor; and (iii) total effects, the sum of direct and indirect effects.

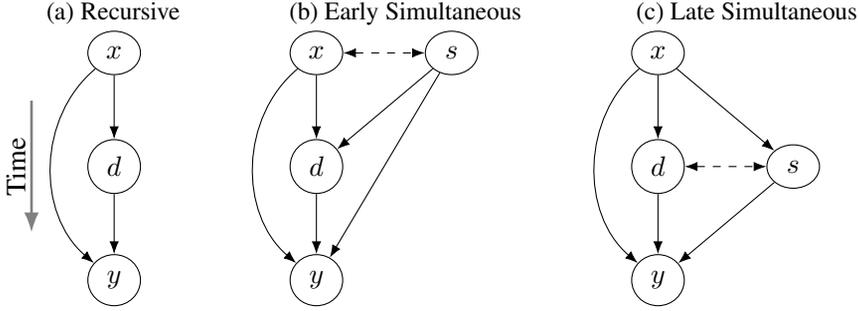
\begin{figure}
    \begin{subfigure}[b]{0.22\linewidth}        
        \caption{Recursive   \textbf{    }}       \begin{tikzpicture}
                \node[state] (x) at (0,3) {$x$};
                \node[state] (d) at (0,1.5) {$d$};
                \node[state] (y) at (0,0) {$y$};
                \path (x) edge (d);
                \path (x) edge[bend right=50] (y);
                \path (d) edge (y);
                \node (w) at (1.1,0) {};
                \node (t) at (-1.1,2.5) {};
                \node (T) at (-1.1 ,0.5) {};
                \path[white] (T) edge node[black, above, el] {Time} (t);
                \path (t) edge[gray, line width=1] (T);
        \end{tikzpicture}
        \label{A}
    \end{subfigure}
    \begin{subfigure}[b]{0.32\linewidth}        
           \caption{Early Simultaneous} 
           \begin{tikzpicture}
               \node[state] (d) at (0,1.5) {$d$};
                \node[state] (x) at (0,3) {$x$};
                \node[state] (s) at (1.8,3) {$s$};
                \node[state] (y) at (0,0) {$y$};
                \path (x) edge (d);
                \path (d) edge (y);
                \path (s) edge (d);
                \path (s) edge (y);
                \path (x) edge[bend right=50] (y);
                \path[bidirected] (x) edge (s);
            \end{tikzpicture}
        \label{B}
    \end{subfigure}
    \begin{subfigure}[b]{0.32\linewidth}        
        \caption{Late Simultaneous}
        \begin{tikzpicture}
            \node[state] (x) at (0,3) {$x$};
            \node[state] (d) at (0,1.5) {$d$};
            \node[state] (s) at (1.8,1.5) {$s$};
            \node[state] (y) at (0,0) {$y$};
            \path (s) edge (y);
            \path (x) edge (d);
            \path (x) edge (s);
            \path (x) edge[bend right=50] (y);
            \path (d) edge (y);
            \path[bidirected] (d) edge (s);
        \end{tikzpicture}
        \label{C}
    \end{subfigure}
    \caption{Three directed-acyclic graphs representing three causal systems: (a) fully identified and recursive; (b) unidentified on account of an earlier-determined simultaneous regressor; and (c) system (b) but with a later-determined simultaneous regressor instead. Arrows indicate the directions of causality of the recursive regressors $x$, $d$,  and simultaneous regressor $s$. The passage of time is shown on the far left.}
    \label{causal}
\end{figure}

To illustrate, consider the linear structural equation representation for the recursive example in Figure 1(a), where $x$ and $d$ are centered and ordered regressors; $y$ is the independent variable; $\beta_{ij}$, $i,j=x,d,y$, are the unobservable population parameters; and $\upsilon_i$ is the residual: \begin{align} 
        \label{eqn:structural_x}
        x &= \upsilon_x, \\
        \label{eqn:structural_d}
        d &= \beta_{xd} x + \upsilon_d,  \\
        \label{eqn:structural_y}
        y &= \beta_{xy} x + \beta_{dy} d + \upsilon_y.
    \end{align}  

\noindent Here, variables $x$ and $d$ are recursive because they occur one at a time in causal order by way of a temporal or experimentally designed separation. Specifically, $x$ is determined before $d$ and then affects $d$ directly through parameter $\beta_{xd}$. Regressor $x$ is also determined before y, acting upon $y$ directly through $\beta_{xy}$. Finally, $x$ exerts an indirect effect on $y$ by way of its influence on $d$, in turn influencing $y$ through $\beta_{dy}$. Together, the total effect of $x$ on $y$ is the sum of its direct and indirect effects, i.e., $(\beta_{xy} + \beta_{xd}\beta_{dy})$. By the chain rule, this total effect is also the partial derivative of $y$ with respect to $x$. 

We can now assemble the total effects of the reduced form system, where residuals $\upsilon_i$, $i = x, d, y$ now serve as regressors:\begin{align}
        \label{eqn:rf0}
        x &= \upsilon_x, \\
        d &= \beta_{xd} \upsilon_x + \upsilon_d,  \\
        \label{eqn:rf}
        y &= (\beta_{xy} + \beta_{xd}\beta_{dy}) \upsilon_x + \beta_{dy} \upsilon_d + \upsilon_y. 
    \end{align}  

\noindent Note that parameter $\beta_{dy}$ is both the direct and total effect of $d$ on $y$, given that $d$ is the last regressor to be determined and influences no other regressors. This can also be shown by the Frisch-Waugh-Lovell decomposition theorem \citep{frisch, lovell}, first described by \citet{yule}, that $\beta_{dy}$ is recoverable from the regression of the independent variable $y$ on residual $\upsilon_d$ and regressor $d$.  

The source of multicollinearity in this recursive system is that each regressor influences those determined later. For instance, when regressors are standardized and no spurious or incidental multicollinearity is present, the covariance between $x$ and $d$ is exactly the direct-effect parameter $\beta_{xd}$. 

For convenience, the entire reduced-form system, including the dependent variable, can also be represented as a matrix decomposition $X = U(A+I)$, with upper-triangular parameter matrix:\begin{equation}
A=
    \begin{bNiceArray}{ccc}[first-row,code-for-first-row=\scriptstyle]
        \\
        0  & \hspace{0.5cm} \beta_{xd} & \hspace{0.5cm} \beta_{xy} + \beta_{xd}\beta_{dy} \\
        0 & \hspace{0.5cm} 0 & \hspace{0.5cm} \beta_{dy} \\
        0 & \hspace{0.5cm} 0 & \hspace{0.5cm} 0 \\
    \end{bNiceArray}
    \label{eqn:matrixA}
\end{equation}

Matrix $A$ also represents the matrix of first-order partial derivatives because $\tfrac{\partial y} {\partial x} = \beta_{xy} + \beta_{xd} \beta_{dy}$. The parameter matrix $A$ appears similar to the Gram-Schmidt coefficient matrix $C$, which we now formalize.

Under the following assumptions, an equivalence between Gram-Schmidt coefficients $c_{ij} \in C$ in (\ref{eqn:matrixC}), $i,j=1,...,K$, and the reduced-form LSEM parameters $a_{ij} \in A$ in (\ref{eqn:matrixA}) can be shown: 

    \begin{tabular}{l l l}
        (a) &  $X \subseteq \mathcal{R}^{[N \times K]}$ is a full-rank, ordered regressor matrix including dependent variable \\
         & \: \: \: \:  $y = x_K$;  \\
        (b)  & $X$ has positive degrees of freedom $N>K$; and  \\
        (c)  & Error vectors have zero conditional means: $E[\upsilon_i|x_j] = 0$, $j=1,...,i-1$. 
   \end{tabular}

\noindent We will denote the sample residual vectors as $u_i \subseteq U$, $i=1,...,K$.

\vspace{2mm}
\noindent \textbf{Theorem 1 -- Equivalence}. Under assumptions (a) - (c), there exists a unique Gram-Schmidt coefficient matrix $C\subseteq \mathcal{R}^{[K \times K]}$ equivalent in expectation to the recursive LSEM reduced-form matrix $A\subseteq \mathcal{R}^{[K \times K]}$
    \begin{align}  
        E[C]=A.
    \end{align} 
    
\vspace{2mm}
\noindent \textbf{Proof.} See online \hyperref[appn]{Appendix}.

\section{ESTIMATION PROPERTIES}

\noindent In the next theorem, we will compare the finite sample properties of Gram-Schmidt coefficients $c_{ij}$ in equations (\ref{eqn:laplace0}) - (\ref{eqn:laplace}) with those estimated by OLS $b_{ij}$. Define: (\textit{i}) $X_{-j}$ as the full regressor set $X$ excluding regressor $x_j$; (\textit{ii}) $X_{+k}$ as the regressor set including the additional variable $x_k$; (\textit{iii}) $X_{<j}$ as the regressor set including ordered regressors $x_i$ up to, but not including, $x_j$, $i=1,...,j-1$; and (iv) $R_{jX_{<j}}^2$ as the coefficient of determination of the regression of $x_j$ on the set of remaining regressors $X_{<j}$. We will assume the regressor order is $i < k < j$ throughout, in which $x_j$ will be the $j^\text{th}$ equation's dependent variable. We make two additional assumptions:\footnote{These complete the classic assumptions of the Gauss-Markov theorem, with homogeneity imposed by the scalar-value parameters in the structural equations (\ref{eqn:structural_x}) - (\ref{eqn:structural_y}).} 

    \begin{tabular}{l l l}
        (d) &  The true underlying model is given by structural equations (\ref{eqn:structural_x}) - (\ref{eqn:structural_y}); and \\
         (e)  & OLS and Gram-Schmidt residuals $\upsilon_{i}$ are free of autocorrelation and heteroscedasticity, \\
         & \: \: \: \:  $E[\upsilon_i\upsilon_i'|\mathbf{X}]=\sigma_i^2\mathbf{I}$.
   \end{tabular}
\vspace{2mm}

\noindent \textbf{Theorem 2 -- Properties.} Under assumptions (a) - (e), the Gram-Schmidt system $X=U(C+I)$ obtains the following estimation properties: 

    \begin{tabular}{l l l}
        (A) & Regressors are orthogonal, $u_i \perp u_j$; \\
        (B)  & Coefficients are stable,  $Cov[c_{ij},c_{ji}] = 0$;  \\
        (C)  & Coefficients are unbiased,  $E[c_{ij}]=a_{ij}$; \\
        (D)  & All information is preserved,  $R_{j \, X_{<j}}^2 = R_{j \, U_{<j}}^2$;  \\
        (E)  & Omitted-variable bias is zero,  $E[c_{ij}|X]=E[c_{ij}|X_{-k}]$;  \\  
        (F)  & Gram-Schmidt variance is lower than OLS,  $V[c_{ij}] \leq V[b_{ij}]$; and\\
        (G)  & Gram-Schmidt variance is lower than OLS under included-irrelevant variables, \\
        & \: \: \: \:  $V[c_{ij}|X_{+k}] \leq V[b_{ij}|X_{+k}]$. 
   \end{tabular}

\vspace{2mm}
\noindent \textbf{Proof.} See online \hyperref[appn]{Appendix}. Confidence intervals of the total effects and model inference are obtained directly from the terminal Gram-Schmidt regression, eliminating the need to reconstruct indirect effects by way of either product-of-coefficients \citep{alwin} or simulation methods.\footnote{\citet{imai10a} propose what they refer to as a nonparametric indirect-effect estimator, calculated by forecasting the independent variable with and without the regressor interaction of interest. Their model relies on parametric estimators of the structural equations. They show their estimator is consistent when all underlying estimators are also consistent. When regressors are present, model errors lack an analytic asymptotic distribution, so confidence intervals must be simulated \citep[p. 59]{imai10b}.}

\section{CAUSALITY}

\noindent To explore the causal interpretation of the Gram-Schmidt coefficients, we relax the homogeneity assumption implicit in the structural model (\ref{eqn:structural_x}) - (\ref{eqn:structural_y}), allowing individuals to exhibit heterogeneous responses to the treatment. We will examine two cases: the classic (late) treatment case in which a treatment $d$ is administered to a group of individuals with predetermined characteristics $x$, and an alternative (early) treatment case in which the treatment is one such characteristic $x$ or a treatment administered before the formation of such characteristics. The early and late cases are respectively represented by treatment variables $x$ and $d$ in Figure 1(a).

In the late-treatment case $d$, we will show that both the convex combination and causal interpretations of the Gram-Schmidt are identical to those of OLS, introduced by \citet{angrist98} and extended by \citet{sloczynski}. In the early-treatment case $x$ by contrast, the Gram-Schmidt approach provides the unbiased average total treatment effect (ATTE). The average total treatment effect on the treated (ATTT) is identical to that in the untreated (ATTU). This makes intuitive sense because early-treatment total effect $c_x$ includes $x$'s direct effect $\beta_{xy}$ as well as its intermediate influence on the formation of later personal characteristics $\beta_{xd} \beta_{dy}$ - which themselves may include associations with various programs and institutions. In effect, if a control group member were to be moved to the early treatment group, such as being black, all regressor characteristics would follow as if they had been a member of the original treatment group.

Two existence and uniqueness assumptions are required for the convex combination interpretation:

    \begin{tabular}{l l l}
        (f) & Expected values $E[y^2]$ and $E[ \|z \|^2]$ are finite, $z = x,d$; \\
        (g) & The covariance matrix of ($x,d$) is nonsingular; and \\
        (h)  & Variances $V[p(z)|m=j]$ are nonzero, where $m = x,d$, $m\neq z$, and $j=0,1$.
   \end{tabular}
   
\noindent Here $p(z)$ is the probability of treatment, defined as the projection $\hat{z}$ from a regression of the treatment variable $z$ on the predetermined characteristics $m$, as shown in equation (\ref{eqn:structural_d}) for variables $d$ on $x$.

It is also useful to define the projection of $y$ on the probability of treatment $p(z)$, given a treatment group $j=0,1$:

\setlength{\abovedisplayskip}{0pt}
\setlength{\belowdisplayshortskip}{0pt}
\begin{align}
        \label{eqn:pry} 
        L[y \; | \; p(z), m = j] = \alpha_j + \gamma_j p(z).
\end{align}  

\noindent The average partial linear effect (APLE) of the treatment on group $j$ is then

\begin{align}
        \label{eqn:aple} 
        c_{zy(APLE,j)} = (\alpha_1 - \alpha_0) + (\gamma_1 - \gamma_0)E[p(z) \; | \; m = j].
\end{align}

A causal interpretation also requires ignorability of the mean and linear probability of a binary treatment variable:

    \begin{tabular}{l l l}
        (i) & $E[y(j) \; | \; z,m] = E[y(j) \; | \; z]$; and \\
        (j) & $E[y(j) \; | \; z] = \alpha_j + \gamma_j p(z)$.  
   \end{tabular}
   
\vspace{2mm}

We can now state the interpretation of the late-treatment Gram-Schmidt (LTGS).

\vspace{2mm}

\noindent \textbf{Lemma 1 -- Late-treatment Gram-Schmidt is identical to OLS.}

\noindent Under assumptions (a) - (h) and a heterogeneous treatment response: 

    \begin{tabular}{l l l}
        (A) & LTGS is a convex combination of partial linear effects, \\
        & \: \: \: \:  $c_{dy} = \omega_1 c_{dy(APLE,1)} + \omega_0  c_{dy(APLE,0)}$; and \\
        (B) & LTGS is a convex combination of ATTT and ATTU, also assuming (i) - (j), \\
        & \: \: \: \:  $c_{dy} = \omega_1 c_{dy(ATTT)} + \omega_0  c_{dy(ATTU)}$.
   \end{tabular}

\vspace{2mm}

\noindent Here, $\omega_j$ are the convex, variance-weighted treatment proportions defined in \citet{sloczynski}.

\vspace{2mm}

\noindent \textbf{Proof.} (A) The late-treatment Gram-Schmidt coefficient is identical to OLS by the Frisch-Waugh-Lovell decomposition theorem \citep{frisch, lovell}, allowing us to apply the result from \citeauthor{sloczynski}'s Theorem 1 and Corollary 1.

\vspace{2mm}

\noindent \textbf{Lemma 2 -- Early-treatment Gram-Schmidt (ETGS) is the ATTE.}

\noindent Under assumptions (a) - (h) and a heterogeneous treatment response: 

    \begin{tabular}{l l l}
        (A) & ETGS is the ATTE, \\
        & \: \: \: \:   $c_{xy} = E[y \; | \; x = 1] -   E[y \; | \; x = 0]$; and \\
        (B) & ETGS ATTT and ATTU are identical.
   \end{tabular}

\vspace{2mm}
\noindent \textbf{Proof.} See online \hyperref[appn]{Appendix}.

Causal identification is also challenged by endogeneity from unobserved regressors, as explored by \citet{imbens94} and many others. Gram-Schmidt properties under omitted variables, included-irrelevant variables, and late-treatment variables all have interesting implications for unobserved regressors.

\section{EXTENDED GRAM-SCHMIDT LEAST SQUARES}

\noindent The Gram-Schmidt transformation is equivalent to the recursive DAG and LSEM when every regressor is recursive, that is, strictly separated by time or when the experimental design is such that no feedback can occur between regressors. However, the latter is an unlikely condition in practice because regressor simultaneity is present in many naturally occurring -- and even some experimentally designed -- data sets, especially when data collection intervals are wide. In the present section, we extend the Gram-Schmidt process to allow a block of simultaneous regressors to replace a single recursive regressor. This preserves temporal ordering but avoids endogeneity bias because the simultaneous regressors within a block are not regressed on one another. Instead, we construct a block-upper-triangular system of coefficients, the final column representing the total effects on the dependent variable from the relevant regressor. Coefficients thus remain consistent with the dependent variable's partial derivatives with respect to the relevant regressor's path-specific effect described by \citet{wright34} and \citet{pearl00}, excluding feedback effects. By omitting only the mutual regressions of the simultaneous regressors, we allow the recovered total effects to exclude only direct feedback effects among simultaneous regressors. Indirect effects of such feedback that may progress through the system are preserved in the total effects of earlier-occurring regressors.

\subsection{Simultaneous Regressors}

\noindent It is worthwhile first to explore the identification problem raised by these simultaneous regressors. It is well known that simultaneous equations, such as supply and demand systems, are not directly identifiable, so parameters cannot be immediately recovered. The same is true for simultaneous regressors in structural and DAG models. Consider then a simplified system with two simultaneous regressors $s$ and $x$:
\setlength{\abovedisplayskip}{10pt}
\setlength{\belowdisplayshortskip}{0pt}
\begin{align}
    \label{eqn:simy0}
    s &= \beta_{xs} x + \upsilon_s,  \\
    x &= \beta_{sx} s + \upsilon_x,  \\
    \label{eqn:simy}
    y &= \beta_{sy} s +\beta_{xy} x +  \upsilon_y.  
\end{align}

\noindent This system is unidentified because neither the simultaneous coefficients nor the residual vectors are observable. 

Partial derivatives can, however, be obtained by way of the Implicit Function Theorem: 
\begin{align*}
\frac{\partial y} {\partial x} = \beta_{xy} + \beta_{sy} \frac{1+\beta_{xs}}{1+\beta_{sx}}.
\end{align*}

\noindent Unfortunately, the partial derivatives here are also functions of the unrecoverable parameters $\beta_{sx}$ and $\beta_{xs}$. We will refer to these unrecoverable parameters as feedback effects. Direct effects $\beta_{iy}$ \textit{are} however fully recoverable because OLS remains unbiased, $E[b_{iy}] = \beta_{iy}$. We will also find a way to recover some feedback information from earlier-occurring regressors.

Multicollinearity is assured by the simultaneity of $x$ and $s$, illustrated by the off-diagonal elements of the variance-covariance matrix $\Sigma_s$ of the standardized regressors, shown here without the influence of additional spurious or incidental multicollinearity: 
\begin{equation}
\Sigma_s = 
\begin{bNiceArray}{cc}[ first-row,code-for-first-row=\scriptstyle ]
\\
 \frac{1 + \beta_{xs}^2}{(1-\beta_{xs} \beta_{sx})^2} 
    & \hspace{0.3cm} \frac{\beta_{xs} + \beta_{sx}}{(1-\beta_{xs} \beta_{sx})^2} \\[9pt] 
    \frac{\beta_{xs} + \beta_{sx}}{(1-\beta_{xs} \beta_{sx})^2} 
    & \hspace{0.3cm} \frac{1 + \beta_{sx}^2}{(1-\beta_{xs} \beta_{sx})^2} \\
\end{bNiceArray}
\label{eqn:sigmas}
\end{equation}

In the next section, we explore the information that can be recovered when a system contains both simultaneous and recursive regressors.

\subsection{Mixed Simultaneous and Recursive Regressors}

\noindent Consider next a system containing both an earlier- and later-determined simultaneous regressor block as illustrated in Figures 1(b) and 1(c), respectively. We will motivate the problem for the earlier-determined simultaneous block case, Fig 1(b), although the properties of the extended method shown here apply to both cases. 

The structural equations assumed in Fig 1(b) are: 
\begin{align}
    s &= \beta_{xs} x + \upsilon_s, \label{eqn:mix_s}\\
    x &= \beta_{sx} s + \upsilon_x, \\
    d &= \beta_{sd} s + \beta_{xd} x + \upsilon_d, \\
    y &= \beta_{sy} s + \beta_{xy} x + \beta_{dy} d + \upsilon_y.\label{eqn:mix_y}
\end{align}

As in our simultaneity example in subsection 5.1 above, $s$ and $x$ here are simultaneous. But $d$ has replaced $y$ as the third regressor in the system, so dependent variable $y$ is now a function of three regressors. We know direct feedback effects are unrecoverable, while direct effects $\beta_{id}$ and $\beta_{iy}$ are recoverable, though OLS will suffer from steep multicollinearity. This can be seen in the covariance between $s$ and $d$, illustrated in the following equation for standardized regressors with no additional spurious multicollinearity: 
\begin{align*}
Cov[s,d] &= \frac{\beta_{sd}(1+\beta_{xs}^2)+ \beta_{xd} (\beta_{xs} + \beta_{sx})}{(1-\beta_{xs}\beta_{sx})^2}.
\end{align*}

As explored next, it will be possible to recover more information than direct effects alone and continue to remove the recursive portion of the multicollinearity.

\subsection{The Gram-Schmidt Extension}

\noindent Our first step is to derive the reduced-form equations in terms of the recursive residual $\upsilon_d$ -- along with simultaneous regressors $x$ and $s$ -- rather than in terms of the residuals as was the case for $x$ and $d$ in the original Gram-Schmidt model ({\ref{eqn:rf0}})-(\ref{eqn:rf}): 
\begin{align*}
s &= s, \nonumber \\
x &= x, \\
d &= \beta_{sd} s + \beta_{xd} x + \upsilon_d, \nonumber \\
y &= (\beta_{sy} + \beta_{sd} \beta_{dy}) s + (\beta_{xy} + \beta_{xd} \beta_{dy}) x + \beta_{dy} \upsilon_d + \upsilon_y \nonumber .
\end{align*}


This system can be specified as $X = U(A+I)$, where $U = [s \: x \: \upsilon_d \upsilon_y]$ and parameter matrix $A$ -- extending (\ref{eqn:matrixA}) -- is now the block-upper-triangular partial derivatives matrix: 
\begin{equation}
    A=
    \begin{bNiceArray}{cccc}[ first-row,code-for-first-row=\scriptstyle]
    \\
    0 & \hspace{0.5cm} 0 & \hspace{0.5cm} \beta_{sd} & \hspace{0.5cm} \beta_{sy} + \beta_{sd}\beta_{dy} \\
    0 & \hspace{0.5cm} 0 & \hspace{0.5cm} \beta_{xd} & \hspace{0.5cm} \beta_{xy} + \beta_{xd}\beta_{dy} \\
    0 & \hspace{0.5cm} 0 & \hspace{0.5cm} 0 & \hspace{0.5cm} \beta_{dy} \\
    0 & \hspace{0.5cm} 0 & \hspace{0.5cm} 0 & \hspace{0.5cm} 0 
    \end{bNiceArray}
    \label{eqn:matrixA2}
\end{equation}

The extended method is then estimated as: 
\begin{align*}
&           && \negthickspace
                \begin{rcases}  
                    s    = s \\
                    x    = x
                \end{rcases}             \textit{        Block 1,} \\
&           && \negthickspace
                \begin{rcases}  
                    d = c_{sd} s + c_{xd} x + u_d 
                \end{rcases}            \textit{        Block 2,} \\
&           &&  \negthickspace
                \begin{rcases}  
                    y = c_{sy} s + c_{xy} x + c_{dy} u_d + u_y 
                \end{rcases}            \textit{        Block 3.} 
\end{align*}

\noindent Block 1 consists of the early simultaneous regressors $x$ and $s$, as in Figure 1(b), while blocks 2 and 3 contain only one dependent variable, similar to the recursive case in Figure 1(a). Block 1 feedback coefficients $b_{sx}$ and $b_{xs}$ are omitted, so $x$ and $s$ are not regressed on one another. However, each dependent variable in blocks 2 and 3 is regressed on all regressors in the blocks previous to it so the indirect effects $b_{sd}$ and $b_{xd}$ are recovered.

Residual regressor matrix $U$ is no longer completely orthogonal. Rather, our extended method removes all covariance between any two blocks, in the present case between the simultaneous and recursive blocks, such that (i) the off-diagonal elements of the mixed variance-covariance partitioned matrix $\Sigma_{sr}$ are zero, for instance $Cov[s,u_d] = 0$; but (ii) the simultaneous covariance matrix $\Sigma_s$ from equation (\ref{eqn:sigmas}) remains in the upper left:
\[
\Sigma_{sr} =
\left[\begin{array}{cc|c}
       \Sigma_s & \hspace{0.4cm} & \hspace{0.2cm} \bigzero  \\ 
\hline
       \bigzero & \hspace{0.4cm} & \hspace{0.2cm} 1  
\end{array}\right],
\]
where $\bigzero$ is a conforming ($2 \times 1$) zeros vector.

The extended method works similarly when a simultaneous regressor block follows a recursive regressor, as in Figure 1(c): 
\begin{align*}
&&& \negthickspace
                \begin{rcases}  
                    x    = x
                \end{rcases}             \textit{        Block 1,} \\
&&& \negthickspace
                \begin{rcases}  
                    s = c_{xs} x + u_s \\
                    d = c_{xd} x + u_d 
                \end{rcases}            \textit{        Block 2,} \\
&&&  \negthickspace
                \begin{rcases}  
                    y = c_{xy} x + c_{sy} u_s + c_{dy} u_d + u_y
                \end{rcases}            \textit{        Block 3.} 
\end{align*}

\noindent Block 2 now contains the two simultaneous regressors $s$ and $d$ in Figure 1(c) that will not be regressed on one another, although each will be regressed on $x$ in block 1. Covariances between pairs of blocks are again removed in the mixed variance-covariance partitioned matrix $\Sigma_{rs}$. For instance $Cov[u_s,u_d] = 0$ and the simultaneous covariance matrix $\Sigma_s$ shifts to the lower right: 
\[
\Sigma_{rs} =
    \left[\begin{array}{ccc|c}
       1 & \hspace{0.5cm} & \hspace{0.5cm} & \hspace{0.1cm} \bigzero  \\ 
        \hline
       \bigzero & \hspace{0.5cm} & \hspace{0.5cm} & \hspace{0.1cm} \Sigma_s  
    \end{array}\right].
\] 

\subsection{Extended Estimation Properties}

\noindent We can now state the properties of our extended Gram-Schmidt least-squares (GSLS) method, which achieves the Theorem 2 properties across blocks rather than across individual regressors. Multicollinearity is eliminated from the recursive regressors, and recursive total effects are recovered. Multicollinearity among simultaneous regressors in a given block will be reduced but not eliminated. Feedback effects recovered, but the direct and intermediate effects are included in the simultaneous regressors' total effects. 

To show this, we revise the orthogonality assumption (a) in Theorems 1 and 2. Let $X \subseteq \mathcal{R}^{[N \times K]}$ be a full-rank matrix of \textit{recursively} ordered regressor blocks $m,n,h=1,...M$, each block $m$ containing one or more regressors with coefficient $c_{i(m)j(n)}$, and block $m$ determined prior to block $n$, $m<n$. The true underlying model in assumption (d) is now represented by block structural equations (\ref{eqn:mix_s})-(\ref{eqn:mix_y}). Throughout, we assume the regressor order to be $i < k < j$ and the block order to be $m < h < n$, with $x_{j(n)}$ the $j^\text{th}$ equation's dependent variable.

With this revision in mind, we can summarize the properties of the GSLS estimator.
\vspace{2mm}

\noindent \textbf{Theorem 3 -- Properties.} Under our revised assumptions (a) - (e) above, the GSLS system $X=U(C+I)$ obtains the following estimation properties:

    \begin{tabular}{l l l}
        (A)  & Regressors are orthogonal across blocks, $ u_{i(m)} \perp u_{j(n)}$; \\
        (B)  & Coefficients are stable across blocks,  $Cov[c_{i(m)j(n)},c_{k(h)j(n)}] = 0$;  \\
        (C)  & Coefficients are unbiased,  $ E[c_{i(m)j(n)}]=a_{i(m)j(n)}$; \\
        (D)  & All information is preserved,  $R_{j(n) \, X_{<j(n)}}^2 = R_{j(n) \, U_{<j(n)}}^2$;  \\
        (E)  & Omitted-variable bias is zero,  $E[c_{i(m)j(n)}|X]=E[c_{i(m)j(n)}|X_{-k(h)}]$;  \\  
        (F)  & GSLS variance is lower than OLS,  $V[c_{i(m)j(n)}] \leq V[b_{i(m)j(n)}]$; and \\
        (G)  & GSLS variance is lower than OLS under included-irrelevant variables, \\
        & \: \: \: \:   $V[c_{i(m)j(n)}|X_{+k(h)}] \leq V[b_{i(m)j(n)}|X_{+k(h)}]$. 
   \end{tabular}

\vspace{2mm}
\noindent \textbf{Proof.} See online \hyperref[appn]{Appendix}. 

\section{Empirical applications}

\noindent We now compare the GSLS estimator to OLS using two empirical examples. Readers may replicate the analysis here with data and software packages available for \citeauthor{R} and \citeauthor{stata} \citep{cross}. 

\subsection{National Supported Work Program}

\noindent First, we extend \citeauthor{sloczynski}'s replication of \citeauthor{lalond86}'s (\citeyear{lalond86}) and \citeauthor{angrist09}'s (\citeyear{angrist09}) analyses (hereafter SLA) of the National Supported Work (NSW) training program's impact on future earnings. Our late-treatment effect matches the impact of the program reported by SLA. We also look at the early-total treatment effect of being black on both inclusion in the work program and future earnings. GSLS extends the discrimination-decomposition framework of \citet{bohren} to a regression context by decomposing the total-effect estimate into a direct and a systemic discrimination component. We will find that black individuals were included in the program at higher rates and experienced lower earnings than non-black individuals, a phenomenon linked to both direct and systemic discrimination.

\subsubsection{Data}

\noindent Prior studies controlled for several strongly interrelated individual characteristics drawn from the Current Population Survey (CPS) and the Panel Study of Income Dynamics (PSID). NSW participants were randomly assigned a training program and a control group. Specific job assignments were however assigned locally with candidate response rates varying across demographic groups. This resulted in an over-representation of black and economically disadvantaged individuals in the treatment group \citep{sloczynski}. To test the extent to which black individuals were treated differently, we aggregate Hispanic and white participants to create a binary black and non-black treatment variable. This will serve as an early treatment because individuals were black before other study characteristics were formed and inclusion in the job program was assigned.

\def\arraystretch{1.15}
\begin{table}[h]
    {
    \caption{Summary statistics of NSW data.}
    \label{tbl:summaryNSW}
    \sisetup{
      input-symbols         = (),
      table-format          = -1.2,
      table-align-text-post = false,
      group-digits          = false
    } 
    \begin{tabular}{
      l
      S[table-format=-1.2]
      S[table-format=-1.2]
      S[table-format=-1.2]
      S[table-format=-1.2]
      S[table-format=-1.2]
      S[table-format=-1.2]
      S[table-format=-1.2]
      S[table-format=-1.2]
     } 
     \hline
     & \multicolumn{2}{c}{Job Program} & &  \multicolumn{2}{c}{Control} \\
     \cline{2-3} \cline{5-6}
     & {} {Mean} & {Std. dev.} && {Mean} & {Std. dev.} \\
     \hline
     Black (proportion)  & 84\% & 0.4 &  & 7\% & 0.3 \\ 
     White (proportion)  & 10\% & 0.3 &  & 85\% & 0.4 \\ 
     Age  & 25.8 & 7.2 &  & 33.2 & 11.0 \\ 
     Education  & 10.4 & 2.0 &  & 12.0 & 2.9 \\ 
     No degree (proportion) & 71\% & 0.5 &  & 30\% & 0.5 \\ 
     Married (proportion)  & 19\% & 0.4 &  & 71\% & 0.5 \\ 
     Annual earnings 1978 (\$1000s)  & 6.3 & 7.9 &  & 14.8 & 9.6 \\ 
    \hline
   \end{tabular} } 
\end{table}

Table \ref{tbl:summaryNSW} summarizes the NSW data for the 185 job program participants and the 15,992 control group individuals. Several regressor means and standard deviations differ materially between the treatment and control groups, suggesting job assignment completions were not random. In the Causal decomposition subsection, we test for the influence of these sample weightings.

\subsubsection{Results}

\noindent 

\vspace{1mm}
\def\arraystretch{1.15}
\begin{table}[h]
    \caption{Direct and total effects of selected regressors on NSW participants' 1978 earnings (\$1000s).}
    \label{tblresultsNSW}
    \sisetup{
      input-symbols         = (),
      table-format          = -1.2,
      table-align-text-post = false,
      group-digits          = false
    } 
    \begin{tabular}{
          l
          S[table-format=-1.2]
          S[table-format=-1.2]
          S[table-format=-1.2]
          S[table-format=-1.2]
          S[table-format=-1.2]
          S[table-format=-1.2]
          S[table-format=-1.2]
          S[table-format=-1.2]
        } 
        \hline
        & \multicolumn{2}{c}{OLS direct effects} & &  \multicolumn{2}{c}{GSLS total effects} \\
        \cline{2-3} \cline{5-6}
        & {Coeff.} &  {S.E.} & & {Coeff.} & {S.E.} \\
        \hline
         Black             &   -2.23 &   (0.28)  &&   -3.74 &   (0.25)\\
         Age               &   0.13   &   (0.01)   &&   0.14 &   (0.01)\\
         Education          &   0.23   &   (0.04)  &&   0.35 &   (0.03)\\
         No degree         &  -1.09  &   (0.23)  &&  -1.29 &  (0.23)\\
         Married           &   3.21  &  (0.19)   &&   3.28 &   (0.19) \\
         Job program       &  -3.47   &   (0.71)  &&   -3.47   &   (0.71)  \\
         \hline 
         $R^2$                   & \multicolumn{2}{c}{0.12} & &  \multicolumn{2}{c}{0.12} \\
        \hline
   \end{tabular} 
\end{table}

\noindent  Table \ref{tblresultsNSW} compares, for selected regressors, the OLS direct effects and GSLS total effects and standard errors.  In both interpretation and expected value the GSLS coefficients differ from OLS. The latter remains unbiased, $E[b_{xy}] = \beta_{xy}$, in the presence of multicollinearity, and its estimates of structural equation (\ref{eqn:structural_y}) represent direct effects. Coefficients are interpreted in the familiar way, namely as an increase or decrease in $y$ resulting from a unit increase in $x$, \textit{ceteris paribus}, assuming all other regressors are fixed or independent. In linear models, this is the partial derivative of $y$ with respect to $x$, ignoring the regressor interrelatedness that induces multicollinearity. 

GSLS estimates provide the total effect, namely the total unit increase (decrease) in $y$ induced by a one-unit increase in $x$, including any indirect effects on $y$ induced or caused by changes in the remaining $x$-dependent regressors. The latter, by the chain rule, is the partial derivative of $y$ with respect to $x$ given all structural relationships in equations (\ref{eqn:structural_x}) - (\ref{eqn:structural_y}), what \citet{wright34} refers to as the path coefficient or \citeauthor{pearl01} as path-switching (\citeyear{pearl01}) or the path-specific effect (\citeyear{pearl00}). GSLS standard errors are lower by the removal of multicollinearity.

The job program's estimated direct effect in the Table \ref{tblresultsNSW} OLS model is identical to its total effect in the GSLS model because, as predicted by Theorem 2, this program is the final regressor in the model. The program's coefficient suggests program participation reduced annual earnings by \$3.47 thousand, slightly larger than the \$3.44 thousand found by \citet[Table 1, Column 1]{sloczynski}, the difference attributable to our aggregation of white and Hispanic participants.

\subsubsection{Race}

\noindent To understand the rise in GSLS Black coefficient magnitude, we look to the discrimination decomposition framework of \citet{bohren}. They define the total expected discrimination function $\Delta(y^0)$ as the sum of direct and systemic discrimination:
\setlength{\abovedisplayskip}{10pt}
\setlength{\belowdisplayshortskip}{0pt}
\begin{align*}
    \Delta(y^0) &= \bar{\tau}(w,y^0) + \delta(b,y^0),
\end{align*}

\noindent where $b$ and $w$ are groups and $y^0$ is the unobservable initial qualification. The direct and systemic components here can be expressed in terms of expected values:
\setlength{\abovedisplayskip}{10pt}
\setlength{\belowdisplayshortskip}{0pt}
\begin{align}
        \label{eqn:dirdiscrim}
        \bar{\tau}(w,y^0) \hspace{0.2cm} &= \hspace{0.2cm} E[A(w; \hspace{0.1cm} S_i) \hspace{0.1cm} - \hspace{0.1cm} A(b; \hspace{0.1cm} S_i) | G_i=w; \hspace{0.1cm} Y^0_i=y^0], \\
        \label{eqn:sysdiscrim}
        \delta(b,y^0)  \hspace{0.2cm}    &= \hspace{0.2cm} E[A(b; \hspace{0.1cm} S_i) | G_i=w; \hspace{0.1cm} Y^0_i=y^0] \hspace{0.1cm} - \hspace{0.1cm} E[A(b; \hspace{0.1cm} S_i) | G_i=b; \hspace{0.1cm} Y^0_i=y^0].
\end{align}  

\noindent Here, $A()$ is the action function and in our case the earnings function; $S_i$ is the set of signals for individual $i$, in our case the individual's race and personal characteristics; $G_i$ is the set of groups ${w,b}$; and $Y^0_i$ is the set of initial qualifications, which we assume to be equal across all participants in the program since, for the early-treatment, it represents their potential for earnings before they are born.

In words, direct discrimination in (\ref{eqn:dirdiscrim}) is the difference in action A, earnings, received by individual $i$, were they to belong to group $w$ rather than group $b$, holding the individual's other signals (background characteristics) $S_i$ and initial qualification $y^0$ constant at the group-$w$ level. This matches the interpretation of OLS results shown in Table \ref{tblresultsNSW}. The direct discrimination estimate suggests earnings are now lower by a significant \$2.23 thousand dollars per year. 

Systemic discrimination in (\ref{eqn:sysdiscrim}) is the expected difference between the earnings of an individual from group $b$ but with background characteristics typical of group $w$, and the same individual's score given group $b$ background characteristics. It represents the cumulative effects of differences in access to such resources as education and employment, which \citeauthor{bohren} categorizes as \textit{technological} sources. Our total discrimination estimate of \$3.74 thousand less per year is significantly greater than the direct effect magnitude, suggesting systemic discrimination is present.

\subsubsection{Causal decomposition}

\noindent The top panel of Table \ref{causalNSW} reports the regression coefficient and its decomposition into the heterogeneous treatment effect on the treated, untreated, and average participants. The probability of being treated $P(d=1)$ and the variance-adjusted statistical weight $\omega_1$ are also listed along with the treatment and control sample sizes and standard errors. 

\vspace{1mm}
\def\arraystretch{1.15}
\begin{table}[h]
    \caption{Causal decomposition of race and job program effects on participants' 1978 earnings (1000s).}
    \label{causalNSW}
    \sisetup{
      input-symbols         = (),
      table-format          = -1.2,
      table-align-text-post = false,
      group-digits          = false
    } 
    \begin{tabular}{
          l
          S[table-format=-1.2]
          S[table-format=-1.2]
          S[table-format=-1.2]
          S[table-format=-1.2]
          S[table-format=-1.2]
          S[table-format=-1.2]
          S[table-format=-1.2]
          S[table-format=-1.2]
        } 
        \hline
        & \multicolumn{2}{c}{OLS early-treatment} & &  \multicolumn{2}{c}{GSLS early-treatment} & &  \multicolumn{2}{c}{GSLS late-treatment} \\
        & \multicolumn{2}{c}{Black} & &  \multicolumn{2}{c}{Black} & &  \multicolumn{2}{c}{Job program} \\
        \cline{2-3} \cline{5-6} \cline{8-9}
        & {Coeff.} &  {S.E.} & & {Coeff.} & {S.E.} & & {Coeff.} & {S.E.} \\
        \hline
         Regression coefficient & -2.23 & (0.28) && -3.74 & (0.26) && -3.47 & (0.71)\\
        \\
        \multicolumn{2}{l}{\textit{Heterogeneous response}}  \\
        ATE / ATTE                   & -2.69 & (0.28) && -3.74 & (0.26) && -6.73 & (1.20)\\
         ATT / ATTT                  & -0.66 & (0.42) && -3.74 & (0.26) && -3.40 & (0.68)\\
         ATU / ATTU                  & -2.87 & (0.29) && -3.74 & (0.26) && -6.77 & (1.21)\\
         \\
         $P(d=1)$               & 0.08 &&& 0.08 &&& 0.01\\
         $\omega_1$             & 0.92 &&& 0.92 &&& 0.98\\
         Obs. treated           & \multicolumn{2}{c}{1,332} & &  \multicolumn{2}{c}{1,332} & &  \multicolumn{2}{c}{185}  \\
         Obs. control           & \multicolumn{2}{c}{14,845} & &  \multicolumn{2}{c}{14,845} & &  \multicolumn{2}{c}{15,992}  \\
         \\
        \hline
         \multicolumn{2}{l}{\textit{Covariate matching}}  \\
         ATE / ATTE                   & -2.45 & (0.38) && -3.57 & (0.38) && -8.05 & (2.05)\\
         ATT / ATTT                   & -2.11 & (0.31) && -3.81 & (0.30) && -3.45 & (0.85)\\
         \\
         Obs. treated           & \multicolumn{2}{c}{16,177} & &  \multicolumn{2}{c}{16,177} & &  \multicolumn{2}{c}{16,177}\\ 
         Obs. control           & \multicolumn{2}{c}{16,177} & &  \multicolumn{2}{c}{16,177} & &  \multicolumn{2}{c}{16,177}\\
        \hline
   \end{tabular} 
\end{table}

For comparison, the bottom panel of Table \ref{causalNSW} provides rebalanced estimates from covariate matching along with rebalanced treatment and control sample sizes. Matching methods rebalance data by selecting a second, untreated individual with characteristics similar to those of each treated individual in the sample. This method selects the pair minimizing the Mahalanobis distance between regressor means and variances in the two groups \citep{dehejia}.\footnote{As desired in the rebalanced model, matched standardized differences (differences in means) are close to zero; and the matched variance ratios are near unity, suggesting that rebalancing reduces the differences between the probability of moments of race, education, and enrollment in the job program despite significant differences between them in the original data. A complete list of rebalanced probability moments is available for both studies in the replication code repository.} 

In the GSLS late-treatment job program model in Column 3, the heterogeneous response estimates match closely to those in \citet[Table 1, Column 1]{sloczynski}, a result of the small magnitude differences resulting from aggregating white and Hispanic participants. The regression coefficient is not statistically different from the heterogeneous ATTT estimate, consistent with the potentially lower regression bias following from the lower probability of treatment $P(d-1)=$ 0.01 and a statistical weight $\omega_1$ close to unity. The smaller ATTE magnitude follows from the larger ATTU estimate along with the convexity restriction from Lemma 1. Finally, the magnitude of the matching ATTE is also larger than the matching ATTT, though not subject to the Lemma 1 restriction.

The total effect of being black is provided in the center column of Table \ref{causalNSW}. Illustrating Lemma 2, the ATTE, ATTT, and ATTU and their standard errors are identical to one another. This can be seen here from the larger total-effect coefficient magnitude, which includes the systemic impacts of being black on such later-formed personal characteristics as grade level, degree completion, and inclusion in the jobs program. Results from rebalancing are similar to the GSLS regression coefficient but with slightly higher standard errors.

The decomposition of the direct effect of being black - in the leftmost data column of Table \ref{causalNSW} - follows a pattern similar to the decomposition in Column 3 but with a much smaller and statistically nonsignificant ATT. This is again partly due to the low probability of treatment $P(d=1)=$ 0.08 and a greater ATU. The magnitude of the matching ATT is also smaller than the matching ATE but by less than the heterogeneity estimate since matching is not subject to the convexity restriction.

Finally, the center column of Table \ref{causalNSW} provides the heterogeneous decomposition of the GSLS early-treatment effect of being black. As suggested by Lemma 2, the GSLS coefficient and heterogeneity ATTE estimate match, and the ATTE is equal to the ATTT. Covariate matching estimates are both within a standard deviation of the regression coefficient.

\subsection{National Longitudinal Survey of Youth}

\noindent Our second illustration examines data from the \cite{bls} National Longitudinal Survey of Youth (NLSY) used by several researchers \citep{korenman, blau, lubotsky} to explore the role of parental income on child reading comprehension test scores while controlling for several strongly interrelated individual and household characteristics. All three studies cite multicollinearity as a leading motivation of their model design, regressor selection, and interpretation of results.

The NLSY began as a survey of 6,283 women and 6,403 men aged 14 to 21 in 1979 and includes a wide range of characteristics, including employment, income, drug use, marriage, education, cognitive assessments, and childbirth. The sample was rebalanced in 1984 to exclude women serving in the military and again in 1990 and 1991 to correct for the survey's original over-weighting of low-income white women. We will test for any remaining sample selection bias at the end of this section. Sampling frequency in the survey was annual through 1994 and biennial thereafter. A second survey, the Child Supplement, was initiated in 1986 to study the children of women included in the 1979 NLSY cohort, biennially recording cognitive achievement and behavior. 

Childhood cognitive development is drawn from the Peabody Reading Comprehension Test. From 1986 to 2014, we include all complete observations of children aged 6-14 who were attempting the test for the first time. We exclude children of mothers dropped from the survey during the 1984, 1990, and 1991 revisions. The final sample includes, from the original NLSY cohort, 6,550 children born to 3,181 mothers. 

\def\arraystretch{1.15}
\begin{table}[h]
    {
    \caption{\textsc{Summary statistics of NLSY data.}}
    \label{tbl:summaryNLSY}
    \sisetup{
      input-symbols         = (),
      table-format          = -1.2,
      table-align-text-post = false,
      group-digits          = false
    } 
    \begin{tabular}{
      l
      S[table-format=-1.2]
      S[table-format=-1.2]
      S[table-format=-1.2]
      S[table-format=-1.2]
      S[table-format=-1.2]
      S[table-format=-1.2]
      S[table-format=-1.2]
      S[table-format=-1.2]
     } 
     \hline
     & \multicolumn{2}{c}{Higher income} & &  \multicolumn{2}{c}{Lower income} \\
     \cline{2-3} \cline{5-6}
     & {} {Mean} & {Std. dev.} && {Mean} & {Std. dev.} \\
     \hline
     Black (proportion)  & 17\% & 0.4 &  & 45\% & 0.5 \\ 
     White (proportion)  & 64\% & 0.5 &  & 31\% & 0.5 \\ 
     Mother's age  & 35.1 & 5.9 &  & 31.8 & 5.7 \\ 
     Mother's education  & 13.6 & 3.2 &  & 11.6 & 2.6 \\ 
     Mother's AFQT (percentile) & 48.2 & 27.0 &  & 23.0 & 21.0 \\ 
     Spouse present (proportion)  & 91\% & 1.3 &  & 33\% & 0.5 \\ 
     Spouse's age & 34.3 & 6.8 &  & 37.0 & 6.3 \\ 
     Spouse's education  & 12.2 & 7.9 &  & 13.7 & 3.7 \\ 
     Child female (proportion)  & 49\% & 0.5 & & 49\% & 0.5 \\ 
     Child's age  & 7.6 & 1.4 &  & 8.0 & 1.6 \\ 
     Family size  & 4.5 & 1.2 &  & 4.2 & 1.7 \\ 
     Family income ($\$$10K)  & 12.3 & 14.7 &  & 2.6 & 1.3 \\ 
     Child's test score (percentile)  & 63.9 & 24.1 &  & 51.7 & 26.4 \\ 
    \hline
   \end{tabular} } 
\end{table}

Table \ref{tbl:summaryNLSY} summarizes the NLSY data by child, broken out into above- and below-median family income levels with 3,275 observations in each category. Several regressor means and standard deviations differ materially between the high- and low-income groups, suggesting the 1984, 1990, and 1991 NLSY rebalancing efforts were insufficient or did not target income specifically. In the Causal decomposition subsection, we test for the impacts of any remaining influence of sample selection bias.

Consistent with all three prior studies of these data, we include annual fixed effects and recursive regressors recorded on the date of test: child's gender and age; logs of family size and income, and mother's race, age, education, and performance on the 1980 Armed Forces Qualification Test (AFQT), a general knowledge exam administered to all participants in the original NLSY cohort. We also control for spousal presence in the household and spousal age and education to compare results with \citet{lubotsky}. Three simultaneous blocks are specified: (i) annual fixed effects; (ii) mother's race, consisting of a nonwhite indicator variable for black or Hispanic (white the omitted variable); and (iii) spousal characteristics, including spouse's residence in the home. All factors are specified in the natural temporal order in which they are assumed to have been determined. For instance, the child's race is defined in the Child Supplement to be the mother's race originally reported in the NLSY. Race was predetermined at the time of the mother's conception, so precedes the mother's age. In turn, both the mother's race and age were determined before her highest grade level was achieved or her AFQT score was recorded. Spousal characteristics, child age and gender, and family size and income are all recorded when the child attempts the reading and comprehension test for the first time, so they may influence the child's test score, but cannot be influenced by it.

\subsubsection{Results}

\noindent Table \ref{tblresultsNLSY} compares, for selected regressors, the OLS direct effects, GSLS total effects, and standard errors. The 2.31 family income coefficient implies a one-percent income rise boosts relative reading comprehension by 2.31 percentile points and is comparable to the range reported by \citet{blau} and \citet{lubotsky}. 

\vspace{1mm}
\def\arraystretch{1.15}
\begin{table}[h]
    \caption{Direct and Total Effects of Selected Regressors on Child's Reading Scores.}
    \label{tblresultsNLSY}
    \sisetup{
      input-symbols         = (),
      table-format          = -1.2,
      table-align-text-post = false,
      group-digits          = false
    } 
    \begin{tabular}{
          l
          S[table-format=-1.2]
          S[table-format=-1.2]
          S[table-format=-1.2]
          S[table-format=-1.2]
          S[table-format=-1.2]
          S[table-format=-1.2]
          S[table-format=-1.2]
          S[table-format=-1.2]
        } 
        \hline
        & \multicolumn{2}{c}{OLS direct effects} & &  \multicolumn{2}{c}{GSLS total effects} \\
        \cline{2-3} \cline{5-6}
        & {Coeff.} &  {S.E.} & & {Coeff.} & {S.E.} \\
        \hline
        Nonwhite      &   -0.40  &   (0.67)  &&   -10.83 &   (0.55)\\
        Mother's age            &   0.06 &   (0.14)   &&   0.02 &   (0.13)\\
         Mother's education          &   0.41 &   (0.11)  &&   1.80 &   (0.09)\\
         Mother's AFQT  &  0.22 &   (0.01)  &&  0.29 &  (0.01)\\
         Child female   & 5.50 & (0.55) &&  6.49 & (0.55)  \\
         Child's age          &   -5.68 &   (0.20)  &&   -5.75 &   (0.20)\\
         Spouse present & 1.92 &  (2.56)   &&   1.31 &   (2.53) \\
         Family size (log)  &   -8.22 &   (0.92)  &&   -8.30 &   (0.92)\\
         Family income (log)       &   2.31 &   (0.74)  &&   2.31 &   (0.74) \\
         \hline 
         $R^2$                   & \multicolumn{2}{c}{0.28} & &  \multicolumn{2}{c}{0.28} \\
        \hline
   \end{tabular} 
\end{table}

\subsubsection{Race}

\noindent Consistent with \citet{blau}, the OLS estimate suggests no significant direct influence of maternal race on child test scores, indicated in the left column of Table \ref{tblresultsNLSY}.\footnote{Race estimates were not reported in \citet{lubotsky}.} This would be expected if test administration and scoring mechanisms did not differ systematically between racial groups. The GSLS model shows race's total effect reduced reading scores by 10.83 percentile points among nonwhite children, relative to white children, significant at the 99.9\% confidence level. The GSLS nonwhite coefficient standard error is lower than OLS by 19\%, illustrating the impacts of removing multicollinearity.\footnote{For a given GSLS regressor, the standard error reduction falls below that indicated by VIF, as the former identifies and removes each regressor's contribution to total model multicollinearity, while the VIF attributes all model multicollinearity to each regressor successively.} 

\begin{table*}[h]
    \caption{Intermediate GSLS step-wise regressions for selected regressors.}
    
    \label{tbl:intermediate results}
    \sisetup{
      input-symbols         = (),
      table-format          = -1.2,
      table-align-text-post = false,
      group-digits          = false
    } 
    \begin{tabular}{
      l
      S[table-format=-1.2]
      S[table-format=-1.2]
      S[table-format=-1.2]
      S[table-format=-1.2]
      S[table-format=-1.2]
      S[table-format=-1.2]
      S[table-format=-1.2]
      S[table-format=-1.2]
    } 
    \hline
     {Dependent variable} & {Mother's} & {Mother's} & {Mother's} & {Spouse}  & {Family} & {Family} & {Child's} \\
      & {age} & {education} & {AFQT} & {present} & {size} & {income} & {score} \\
     \hline 
         Nonwhite   & -1.82 & -1.11 & -29.78 & -0.32 & -0.01 &  -0.33 & -10.83 \\

          & (0.05) & (0.07) & (0.49) & (0.01) & (0.01) & (0.01) & (0.55)  \\  
         
         Mother's age    && 0.10 & 2.05 & 0.01 & 0.003 & 0.02  & 0.02 \\

          && (0.03) & (0.13) & (0.002) & (0.002) & (0.002) & (0.13) \\
         
         Mother's education    &&& 3.44 & 0.01 & -0.01 & 0.04 &   1.80 \\

         &&& (0.67)	 & (0.002)	 & (0.002)    & (0.003) & (0.09) \\
         
         Mother's AFQT  &&&& 0.004 & 0.001 & 0.005 & 0.29 \\
         
          &&&& (0.00)	& (0.00) & (0.00) & (0.01) \\
        
         Spouse present  &&&&& 0.14 & 0.40 & 1.31 \\

         &&&&& (0.04) & (0.05) & (2.53)  \\  

         Family size (log)  &&&&&&  -0.03 & -8.30 \\

         &&&&&& (0.01) & (0.92) \\  

         Income (log) &&&&&&& 2.31 \\
         
         &&&&&&& (0.74) \\

         \hline
         
         $R^2$  & 0.88 & 0.11 & 0.48 & 0.15 &  0.17 & 0.46  & 0.28 \\
        \hline
   \end{tabular} 
\end{table*}

Systemic discrimination is represented by the negative and significant GSLS total effect for race, significantly greater in magnitude than the direct effect. We can use GSLS to further decompose the systemic effect into its channel-specific mechanisms. Table \ref{tbl:intermediate results} provides the intermediate-stage GSLS regression results for selected regressors. Race's total effect on education can be seen in the second estimates column showing the regression of maternal school-grade completion on maternal race and age. Here, the total race effect on education was 1.11 fewer grades completed on average by nonwhite mothers than white mothers. In turn, each additional grade completed by the mother raised the child's test score by 1.80 percentile points, as indicated in the last column of Table \ref{tbl:intermediate results}. Thus, we have a 2.0 percentile point (1.11 $\times$ 1.80) differential between nonwhite and white children attributable to maternal education. This indirect education channel constituted approximately 18\% of the 10.8-percentile total race effect.

\subsubsection{Causal decomposition}

\noindent Table \ref{causalNLSY} tests for heterogeneity and sample weighting bias. The GSLS late-treatment income effect in Column 3 shows an ATTE estimate of 1.91, lower than the regression coefficient of 2.31. This downward bias results from the higher probability of treatment $P(d=1)=$ 0.50 and a similar variance-adjusted statistical weight $\omega_1$. The high ATTT is offset by a nonsignificant ATTU close to zero.

The OLS early-treatment nonwhite effect estimates are nonsignificant for every heterogeneous response and covariate matching estimate, suggesting there is no evidence of direct discrimination in test administrations. 

The GSLS early-treatment coefficient in the center column again matches the heterogeneous responses exactly. The rebalanced sample matching estimates are - though slightly greater in magnitude - not statistically significantly different from the regression coefficient.

\vspace{1mm}
\def\arraystretch{1.15}
\begin{table}[h]
    \caption{Causal decomposition of race and income effects on children's reading scores (test percentile points).}
    \label{causalNLSY}
    \sisetup{
      input-symbols         = (),
      table-format          = -1.2,
      table-align-text-post = false,
      group-digits          = false
    } 
    \begin{tabular}{
          l
          S[table-format=-1.2]
          S[table-format=-1.2]
          S[table-format=-1.2]
          S[table-format=-1.2]
          S[table-format=-1.2]
          S[table-format=-1.2]
          S[table-format=-1.2]
          S[table-format=-1.2]
        } 
        \hline
        & \multicolumn{2}{c}{OLS early-treatment} & &  \multicolumn{2}{c}{GSLS early-treatment} & &  \multicolumn{2}{c}{GSLS late-treatment} \\
        & \multicolumn{2}{c}{Nonwhite} & &  \multicolumn{2}{c}{Nonwhite} & &  \multicolumn{2}{c}{Higher income} \\
        \cline{2-3} \cline{5-6} \cline{8-9}
        & {Coeff.} &  {S.E.} & & {Coeff.} & {S.E.} & & {Coeff.} & {S.E.} \\
        \hline
         Regression coefficient & -0.40 & (0.67) && -10.83 & (0.55) && 2.31 & (0.74)\\
         \\
         \multicolumn{2}{l}{\textit{Heterogeneous response}}  \\
         ATE / ATTE                    & -0.37 & (0.67) && -10.83 & (0.55) && 1.91 & (0.75)\\
         ATT / ATTT                   & -1.02 & (0.82) && -10.83 & (0.55) && 4.06 & (0.86)\\
         ATU / ATTU                   & 0.34 & (0.81) && -10.83 & (0.55) && -0.23 & (1.03)\\
         \\
         $P(d=1)$               & 0.52 &&& 0.52 &&& 0.50\\
         $\omega_1$             & 0.55 &&& 0.55 &&& 0.59\\
         Obs. treated           & \multicolumn{2}{c}{3,427} & &  \multicolumn{2}{c}{3,427} & &  \multicolumn{2}{c}{3,275}  \\
         Obs. control           & \multicolumn{2}{c}{3,123} & &  \multicolumn{2}{c}{3,123} & &  \multicolumn{2}{c}{3,275}  \\
         \\
       \hline
          \multicolumn{2}{l}{\textit{Covariate matching}}  \\
         ATE / ATTE                   & -1.32 & (0.81) && -11.74 & (0.74) && 1.67 & (1.03)\\
         ATT / ATTT                   & -0.12 & (0.98) && -11.49 & (0.86) && 1.49 & (1.24)\\
         \\
         Obs. treated           & \multicolumn{2}{c}{6,550} & &  \multicolumn{2}{c}{6,550} & &  \multicolumn{2}{c}{6,550}\\ 
         Obs. control           & \multicolumn{2}{c}{6,550} & &  \multicolumn{2}{c}{6,550} & &  \multicolumn{2}{c}{6,550}\\
        \hline
   \end{tabular} 
\end{table}

\section{CONCLUSION}

\noindent Multicollinearity has posed a statistical challenge for over 200 years. We have shown that when coefficients are linearly separable through temporal order or experimental design, the Gram-Schmidt process removes multicollinearity and produces economically interpretable estimates, equivalent in expected value to total effects from a recursive Linear System of Equations Model or Directed Acyclic Graph. Total effects and statistical inference follow directly from the regression, eliminating the need for ex-post simulation or product-of-coefficients methods. The coefficients are unbiased estimates of the partial derivatives, invariant to the omission of later-determined regressors, with standard errors lower than in OLS. For early treatments, Gram-Schmidt returns the average total treatment effect on the treated even when the treatment response is heterogeneous.

We have exploited these properties to extend the Gram-Schmidt approach to recover causal effects from a previously unidentified mixed system of recursive and simultaneous regressors. The extended approach removes multicollinearity between recursive and simultaneous regressors, allowing unbiased estimates of the total effects that are free from multicollinearity. 

Using a mix of temporally ordered and simultaneous individual characteristics, we have illustrated extended Gram-Schmidt least squares by applying it to previous studies using the National Supported Work Program and National Longitudinal Survey of Youth. GSLS total effects tended to be greater in magnitude than OLS direct effects, especially among early-determined or inter-related regressors. The approach expanded our interpretation of discrimination impacts reported in earlier studies and lowered standard errors by removing multicollinearity. 


\newpage

\begin{appendix}


\noindent \textsc{\normalsize{APPENDIX}}\label{appn}

\vspace{3mm}
\noindent \textbf{Proof of Theorem 1}:  Existence, uniqueness, and orthogonality are shown by \citet[pp. 57-59]{wong}. 

Coefficients $a_{ij}$ of reduced-form parameter matrix A in (\ref{eqn:matrixA}) can be expressed as the recursive sequence \begin{equation*}
    a_{ij}  = \beta_{ij}+\sum_{n=i+1}^{j-1} a_{in}\beta_{nj},
\end{equation*}

\noindent for $i<j$, and zero otherwise.

Define vector $k_i=u_i'/(u_i'u_i)$ and note that $k_i'x_i=k_i'u_i=1$.

Now, equivalence in expectation between Gram-Schmidt coefficient matrix $C$ in (\ref{eqn:matrixC}) and the reduced-form true, underlying matrix A in (\ref{eqn:matrixA}), $E[c_{ij}] = a_{ij}$, can be shown recursively, since by assumptions (a)-(c) and Theorem 2(A) for each coefficient: \begin{align}
    E[c_{ij}]   &= E[k_i' \, x_j], \\
                &= E[k_i'(\sum_{n=1}^{j-1} x_n \, \beta_{nj} + \upsilon_j)], \\
                &= E[\sum_{n=i}^{j-1} k_i' \, x_n \, \beta_{nj} + k_i' \, \upsilon_j], \\
                \label{unbiasedresult0}
                &= \beta_{ij}+\sum_{n=i+1}^{j-1} E[c_{in}] \beta_{nj} , \\
                \label{unbiasedresult1}
                &= \beta_{ij}+\sum_{n=i+1}^{j-1} a_{in} \beta_{nj},
\end{align}

\noindent for $i<j$, and zero otherwise. Line (\ref{unbiasedresult1}) holds because $E[c_{ij}] = a_{ij}$ follows directly from (\ref{unbiasedresult0}) for $j=i+1$, proving the $j=i+2$ case, and so forth through $j=K$.
\\
\\
\noindent \textbf{Proof of Theorem 2}: 
(A) The orthogonality of the regressors is shown by \citet[pp. 57-59]{wong}. 

(B) Stability of coefficients follows directly from their orthogonality (A) since coefficient covariance 
\begin{equation*}
    Cov[c_{ik},c_{jk}]  = \sigma_{\upsilon_k}^2 \frac{-r_{u_{i}, u_{j}} }{1-r_{u_{i}, u_{j}}^2}
\end{equation*}

\noindent is linear in regressor correlation $r_{u_i,u_j}=u_i'
u_j / \surd(u_i'u_i \; u_j'u_j)$ and zero for any recursive regressor pair.

(C) Unbiasedness follows from the equivalence of conditional expectations in Theorem 1. 

(D) Information preservation holds since the $R^2$ is a linear function of squared residual $u_j$, which in turn is preserved in the reduced form \begin{equation*}
    R_{jX_{<j}}^2 = 1-\tfrac{u_j'u_j}{x_j'x_j}=R_{jU_{<j}}^2.
\end{equation*}

(E) Zero omitted-variable bias follows directly for the later-occurring regressors $k>i$ since regressors are orthogonal by 2(A) and coefficients are unbiased by 2(C).

(F) Variances of Gram-Schmidt coefficients are lower than in OLS. Define (i) $X_{-i}$ as the regressor set $X$ excluding regressor $x_i$; (ii) $X_{<i}$ as the regressor set excluding later-determined regressors $x_i,...,x_K$; (iii) $B \subseteq\mathcal{R}^{[j-1 \times 1]}$ as the OLS coefficient vector; and (iv) $C_i \subseteq\mathcal{R}^{[K \times 1]}$ as the $i^\text{th}$ Gram-Schmidt coefficient vector. Consider the non-trivial case when $\beta_{ij} \neq 0$. 

\vspace{2mm}

\noindent \textbf{Lemma 3.}
    The coefficient of determination weakly declines as regressors are excluded: \begin{align*}
        R_{x_{i} \; X_{<i}}^2 \leq R_{x_{i} \; X_{-i}}^2 .
    \end{align*}

\vspace{2mm}

\noindent This lemma is true for any regressor order, though for convenience, we specify order-of-exclusion in terms of regressors determined after $x_i$. The proof is standard and omitted here.

The variance of OLS coefficient $b_{ij}$ is a function of the $i^{\text{th}}$ diagonal element of the variance-covariance matrix, which can be expressed, by virtue of the \citeauthor{schur}, in terms of the $j^{th}$ regression coefficient of determination: \begin{align*}
    Var[b_{ij}] &= \sigma_{\upsilon_{u_j}} (X'X)_{ii}^{-1} , \\
                &= \sigma_{\upsilon_{u_j}} (x_i'x_i - B_{-i}'(X_{-i}'X_{-i})B_{-i})^{-1}, \\
                &= \sigma_{\upsilon_{u_j}} (x_i'x_i (1-R_{x_{i} \; X_{-i}}^2))^{-1}.
\end{align*}

The variance of Gram-Schmidt coefficient $c_{ij}$ is by 2(A) a function of the residual $u_i$, which by 2(A) and Lemma 3 is: \begin{align*}
    Var[c_{ij}] &= \sigma_{\upsilon_{u_j}} (u_i'u_i)^{-1} , \\
                &= \sigma_{\upsilon_{u_j}} (x_i'x_i - C_i'(X'X)C_i)^{-1}, \\
                &= \sigma_{\upsilon_{u_j}} (x_i'x_i (1-R_{x_{i} \; X_{<i}}^2))^{-1}, \\
                &\leq \sigma_{\upsilon_{u_j}} (x_i'x_i (1-R_{x_{i} \; X_{-i}}^2))^{-1}.
\end{align*}

\noindent The result holds with strict inequality for $i=1,...j-2$ and with strict equality for $i=j-1$ because the excluded regressor set is identical between the terminal Gram-Schmidt and the OLS coefficient.

(G) Gram-Schmidt coefficients have lower variance than OLS when a later-determined irrelevant variable $x_k$ is included because the recursive efficiency gain (F) from Lemma 3 is preserved: \begin{align*}
    R_{x_i \; X_{<i+k}}^2 \leq R_{x_i \; X_{-i+k}}^2.
\end{align*}

\vspace{2mm}

\noindent \textbf{Proof of Lemma 2:} (A) Early-treatment Gram-Schmidt is the ATTE. Recall from Lemma 1(A) the GSLS coefficient $c_{xy}$ is a convex combination of partial linear effects, 
\begin{align}
    c_{xy} = \omega_1 c_{xy(APLE,1)} + \omega_0  c_{xy(APLE,0)}.
\end{align} 

The early-treatment model is given by reduced form equation (\ref{eqn:rf}) expressed below in terms of $x$:

\begin{align}
        y &= (\beta_{xy} + \beta_{xd}\beta_{dy}) x + \beta_{dy} \upsilon_d + \upsilon_y,
    \end{align}  

\noindent where $c_{xy}$ is the coefficient estimate of $\beta_{xy} + \beta_{xd}\beta_{dy}$. The linear probability of treatment $p(\upsilon_d)$ is defined as the projection

\begin{align}
        L[x \; | \; \upsilon_d] = \alpha_p + \beta_p \upsilon_d.
    \end{align}  

\noindent Because $x$ and $\upsilon_d$ are orthogonal by Theorem 2(A), $\beta_p = 0$ and the projection is just a regression on the ones vector. The remaining intercept term $\alpha_p$ is then the unconditional sample mean $E[x]$.

Incorporating this result, the projection of $y$ on the probability of treatment from (\ref{eqn:pry}),

\begin{align}
        L[y \; | \; p(\upsilon_d), x = 1]   &= \alpha_1 + \gamma_1
        p(\upsilon_d), \\
                                            &= \alpha_1 + \gamma_1 E[x], \\
                                            &= E[y \; | \; x=1],
\end{align}  

\noindent is now the regression of $y$ on a constant and equal to the mean outcome when $x$ is positive.

The average partial linear effect of $x$ on $y$ from (\ref{eqn:aple}) can now be simplified 

\begin{align}
        c_{xy(APLE,1)} &= (\alpha_1 - \alpha_0) + (\gamma_1 - \gamma_0)E[p(\upsilon_d) \; | \; x=1], \\
                        &= (\alpha_1 - \alpha_0) + (\gamma_1 - \gamma_0)E[x],\\
                        &= (\alpha_1 + \gamma_1 E[x]) - (\alpha_0 + \gamma_0 E[x]), \\
                        &=  E[y \; | \; x=1] -  E[y \; | \; x=0].
    \end{align}  

\noindent The same is true for $c_{xy(APLE,0)}$ and thus for the convex combination of the two, completing the proof.

\vspace{2mm}

(B) Early-treatment Gram-Schmidt ATTT equals ATTU. 

From Lemma 1(B) and the identity of the Graham-Schmidt coefficient $c_{xy}$ as total effect $(\beta_{xy} + \beta_{xd}\beta_{dy})$, late-treatment Graham-Schmidt is the convex combination of the average total treatment effect on the treated and untreated, 
\begin{align}
    c_{xy} = \omega_1 c_{xy(ATTT)} + \omega_0  c_{xy(ATTU)}.
\end{align} 

From Lemma 2(A) we have $c_{xy(APLE,0)} = c_{xy(APLE,1)} =$ ATTE. Assumptions (i) and (j) and Lemma 1(B) imply $c_{xy(ATTT)}=c_{xy(APLE,1)}$ and $c_{xy(ATTU)}=c_{xy(APLE,0)}$, completing the proof.

\vspace{2mm}

\noindent \textbf{Proof of Theorem 3}: (A) GSLS regressors are orthogonal across blocks.
Define $U_{(m)}$ as the residual block $m<n$ and $U_{<j(-m)}$ the set of all residuals excluding later-determined residuals $u_j,...,u_K$ as well as residual block $U_{(m)}$. Theorem 3(A) then follows from Theorem 2(A) by the inclusion of residual block $U_{(m)}$ in the OLS regression of the first regressor, $x_{j(n)}$, in block $n$: \begin{equation*}
    x_{j(n)} = U_{<j(-m)} C_{<j(-m)} + U_{(m)} C_{(m)} + u_{j(n)}.  
\end{equation*}

\noindent This holds for all regressors in block $n$, since we may reorder any simultaneous regressor arbitrarily to be the first regressor in the block.

(B) Stability of coefficients across blocks follows directly from their orthogonality across blocks (A) and their stability from Theorem 2(B) \textit{mutatis mutandis}.

(C) Unbiasedness. Coefficients $a_{i(m)j(n)}$ of reduced-form parameter matrix A in (\ref{eqn:matrixA2}) can be expressed as the recursive sequence \begin{equation*}
      a_{i(m)j(n)} = \beta_{i(m)j(n)}+\sum_{h=i+1}^{j-1} a_{i(m)h(k)}\beta_{h(k)j(n)},
\end{equation*}
\noindent for $m < n$, and zero otherwise.

Define:  (i) regressor sub-matrix $X_{(m)}$ to include all  $s$ simultaneous regressors in block $m$, $s=1,...,K$; (ii) coefficient sub-vector $c_{(m)j(n)}$ to be the corresponding coefficients in the $j^{th}$ GSLS regression; and (iii) matrix $k_{(m)}=(U_{(m)}'U_{(m)})^{-1}U_{(m)}$. Note $k_{(m)}'X_{(m)}=k_{(m)}'U_{(m)}=I$, the $[s \times s]$ identity matrix. 

Equivalence in expectation between GSLS coefficient matrix $C$ and the reduced-form, true, underlying matrix A in (\ref{eqn:matrixA2}), namely $E[c_{(m)j(n)}] = a_{(m)j(n)}$, can now be shown recursively since by the revised assumptions (a)-(c) and Theorem 3(A), we have for each coefficient: \begin{align}
E[c_{(m)j(n)}]  &= E[k_{(m)}' \, x_{j(n)}], \\
                &= E[k_{(m)}'(\sum_{h=1}^{j-1} x_{h(k)} \, \beta_{h(k)j(n)} + \upsilon_{j(n)})], \\
                &= E[k_{(m)}' \, X_{(m)} \, \beta_{(m)j(n)} \\
                &\:\:\:\:\:\: +\sum_{h=i+s}^{j-1} k_{(m)}' \, x_{h(k)} \, \beta_{h(k)j(n)} + k_{(m)}'  \upsilon_{j(n)}], \\
                \label{unbiasedresult30}
                &= \beta_{(m)j(n)}+\sum_{h=i+s}^{j-1} E[c_{(m)h(k)}] \beta_{h(k)j(n)} , \\
                \label{unbiasedresult31}
                &= \beta_{(m)j(n)}+\sum_{h=i+s}^{j-1} a_{(m)h(k)} \beta_{h(k)j(n)},
\end{align}

\noindent for $m<n$, and zero otherwise. Line (\ref{unbiasedresult31}) holds because $E[c_{(m)h(k)}] = a_{(m)h(k)}$ is shown by equation (\ref{unbiasedresult30}) for $j=i+1$, in turn proving the $j=i+2$ case and so forth until $j=K$.

(D) Information preservation follows from the preservation of reduced-form residuals in Theorem 2(D) because we may arbitrarily order regressors in the simultaneous block $(n)$ such that $x_j$ is the first regressor in the block.

(E) Zero omitted-variable bias follows directly for regressors in later-occurring blocks $k(h)>i(m)$ since regressors are orthogonal across blocks 3(A) and coefficients are unbiased by 3(C).

(F) GSLS coefficients have lower variance than OLS. Define (i) $X_{-i(m)}$ as the regressor set $X$ excluding regressor $x_{i(m)}$; (ii) $X_{ < i(m)}$ as the regressor set excluding regressors in later-determined blocks $m+1,...,M$; (iii) $B \subseteq\mathcal{R}^{[j-1 \times 1]}$ as the OLS coefficient vector; and (iv) $C_{i(m)} \subseteq\mathcal{R}^{[K \times 1]}$ the $i^\text{th}$ as the GSLS coefficient vector. Consider the non-trivial case when $\beta_{i(m)j(n)} \neq 0$ and order block $m$ so that $x_{i(m)}$ is the last regressor in the block.

The variance of OLS coefficient $b_{i(m)j(n)}$ can now be expressed in terms of the coefficient of determination by the Schur Complement: \begin{equation*}
    Var[b_{i(m)j(n)}] = \sigma_{\upsilon_{u_{j(n)}}} (x_{i(m)}'x_{i(m)} (1-R_{x_{i(m)} \; X_{-i(m)}}^2))^{-1}.
\end{equation*}

\noindent The variance of GSLS coefficient $c_{i(m)j(n)}$ is a function of the residual $u_i$ by virtue of 3(A), in which by 3(A) and Lemma 3, 
\begin{align*}
    Var[c_{i(m)j(n)}] &= \sigma_{\upsilon_{u_{j(n)}}} (u_{i(m)}'u_{i(m)})^{-1} , \\
                &= \sigma_{\upsilon_{u_{j(n)}}} (x_{i(m)}'x_{i(m)} (1-R_{x_{i(m)} \; X_{<i(m)}}^2))^{-1}, \\
                &\leq \sigma_{\upsilon_{u_{j(n)}}} (x_{i(m)}'x_{i(m)} (1-R_{x_{i(m)} \; X_{-i(m)}}^2))^{-1}.
\end{align*}

\noindent The result holds with strict inequality for $m=1,...M-1$ and strict equality for $m=M$ because, in the terminal block, the excluded regressor set in GSLS is identical to that in OLS.

(G) GSLS coefficients $c_{i(m)j(n)}$ have lower variance than OLS coefficients $b_{i(m)j(n)}$ when irrelevant variable $x_{k(h)}$ is included in a later-occurring block $m<h$ because recursive efficiency gain 2(F) from Lemma 3 is preserved: 
\begin{align*}
    R_{x_{i(m)} \; X_{<i(m)+k(h)}}^2 \leq R_{x_{i(m)} \; X_{-i(m)+k(h)}}^2.
\end{align*}

\end{appendix}

\end{document}